\documentclass[12pt]{article}   
\usepackage[pdftex]{graphicx}
\usepackage[pdftex]{color}
\usepackage{times}
\title{TRI-BN-15-03: Linac Envelope Optics}
\author{Rick Baartman, TRIUMF}
\date{March, 2015}
\begin{document}
\maketitle
\raggedright
\begin{abstract}
I develop the formalism that allows calculation of beam envelopes through a linear accelerator given its on-axis electric field. Space charge can naturally be added using Sacherer\cite{sacherer1970rms} formalism. A complicating feature is that the reference particle's energy-time coordinates are not known a priori. Since first order matrix formalism applies to deviations from the reference particle, this means the reference particle's time and energy must be calculated simultaneously with the beam envelope and transfer matrix. The code {\tt TRANSOPTR}\cite{dejong1983first} is used to track envelopes for general elements whose infinitesimal transfer matrices are known, and in the presence of space charge. Incorporation of the linac algorithm into {\tt TRANSOPTR} is described, and some examples given. 
\end{abstract}

\section{Introduction}
For beamline design, matching, etc., even in the case of strong space charge forces, it is often sufficient to calculate the evolution of size parameters such as the rms width and bunch length. The full set of such parameters is the 6-dimensional ``$\sigma$-matrix''; $x,P_x,y,P_y,z,P_z$. Conventionally, the momenta are divided by the reference particle's momentum, making them $x',y',\Delta P/P$\footnote{Throughout this note, primes denote $d/ds$}. Parameter 5 is technically time, but converted to distance (bunch length) by multiplying by the speed $\beta_0c$ of the reference particle. Similarly, parameter 6 is $\Delta E$, converted to $P_z$ by dividing by $\beta_0c$. The $\sigma$-matrix has 21 parameters, as it is symmetric: 6 rms sizes on the diagonal, 15 correlation parameters. 

The mathematical fomalism for this technique, including space charge, was established by Frank Sacherer\cite{sacherer1970rms}. The space charge forces depend crucially upon the bunch dimensions in configuration space, so it is important that these be tracked. In other words, it is insufficient to use a formalism that first integrates the equations of motion to derive the transfer matrices, and apply space charge effects afterwards. Some implementations such as {\tt TRANSPORT} and {\tt TRACE3D} divide standard elements into (hopefully sufficiently short) segments interleaved with space charge ``lenses''. This is crude, approximate and non-adaptive.

{\tt TRANSOPTR}\cite{dejong1983first} uses the envelope formalism, but did not until now include the case of beam high intensity bunches being accelerated with a RF accelerator. This case is of particular importance for modeling the elecron linear accelerator. Short accelerator gaps can often be sufficiently well modelled as infinitesimal i.e.\ as a thin longitudinal lens with also a transverse thin lens focusing component. But this is insufficient for extended RF devices and as well misses time-of-flight effects when the change in velocity of the reference particle is significant. Extended DC longitudinal fields have already been added to {\tt TRANSOPTR}\cite{baartman2010acc}; the present study aims at adding the AC case. In the DC case, the reference particle's longitudinal coordinates are inferred from the local potential. Not so the AC case; the time and energy versus longitudinal location must be tracked in a separate calculation and then the equations of first order deviations applied to find the transfer matrices and local envelope optics.

\section{Theory}
This follows closely from the previous note on DC axially-symmetric longitudinal fields\cite{baartman2010acc}.
\subsection{Hamiltonian}
With the distance along the reference trajectory $s$ as the independent variable, the Hamiltonian is
\begin{equation} \label{ham1}
H(x,P_x,y,P_y,t,E;s)=-qA_s-\sqrt{\left(\frac{E-q\Phi}{c}\right)^2-m^2c^2-(P_x-qA_x)^2-(P_y-qA_y)^2}
\end{equation} 

The case of RF axially-symmetric electric field can be handled entirely with no electric potential ($\Phi=0$), and time-varying vector potential. This has been presented a number of times in the past (e.g. Chambers\cite{chambers1968particle}), but we are interested in the following more experimentally-useful case: The electric field along the axis ${\cal{E}}(s)$ has been measured and is therefore known, and the geometry is exactly axially symmetric. Rob Ryne\,\cite{rynerfgap1991} has treated this case, and we use his vector potential $\vec{A}(x,y,s,t)$ directly.
\begin{eqnarray}\label{ax}
 A_x&=&\frac{{\cal{E}}'(s)}{2}\frac{\sin(\omega t+\theta)}{\omega}\ x\\\label{ay}
 A_y&=&\frac{{\cal{E}}'(s)}{2}\frac{\sin(\omega t+\theta)}{\omega}\ y\\\label{as}
 A_s&=&\left(-{\cal{E}}(s)+\frac{x^2+y^2}{4}\left[{\cal{E}}''(s)+\frac{\omega^2}{c^2}{\cal{E}}(s)\right]\right) \frac{\sin(\omega t+\theta)}{\omega}
\end{eqnarray}
This is Coulomb/Lorenz gauge, satisfies Maxwell equations to second order in transverse coordinates, gives correct on-axis  $\vec{\cal{E}}=-\partial\vec{A}/\partial t={\cal{E}}\cos(\omega t+\theta)$.

A priori, we do not know the reference particle's energy and time coordinates. We need these in order to expand about them. They can be found from the equations of motion for $x=y=P_x=P_y=0$:
\begin{eqnarray}
 \frac{dE}{ds}&=&\frac{\partial H}{\partial t}=q{\cal{E}}\cos(\omega t+\theta)\\
 \frac{dt}{ds}&=&-\frac{\partial H}{\partial E}=\frac{E}{c^2P}=\frac{1}{\beta_0c}
\end{eqnarray}
(From here on, I drop the $0$ subscript: $\beta$ and $\gamma$ are implicitly assumed to be the relativistic parameters of the reference particle.)

These are solved first and give the functions $E_0(s)$ and $t_0(s)$ about which $t$ and $E$ are expanded: $E=E_0+\Delta E$, $t=t_0+\Delta t$. So we transform the canonical variables $t$ and $-E$ to $(\Delta t,-\Delta E)$, using as generating function
\begin{equation} G=-\left(t-\int\frac{ds}{\beta(s) c}\right)(\Delta E+E_0)\end{equation} 
(Check: $\frac{\partial G}{\partial t}=-E$, $\frac{\partial G}{\partial(-\Delta E)}=\Delta t$.)
The Hamiltonian gets the added terms \[\frac{\partial G}{\partial s}=\frac{\Delta E+E_0(s)}{\beta(s)c}-\Delta tE_0'(s).\] Then expanding the square root, we get:
\begin{equation} \label{hamdt}
H_{\Delta t}=\left(\frac{E_0}{\beta c}-P_0\right)-qA_s-\Delta tE_0'(s)+\frac{(\Delta E)^2}{2\beta^3\gamma^3mc^3}+\frac{(P_x-qA_x)^2+(P_y-qA_y)^2}{2P}
\end{equation} 

In expanding $P_x-qA_x,P_y-qA_y$, the time dependence disappears because it is higher order:
\begin{equation}
 (P_x-qA_x)^2=P_x^2-q{\cal{E}}'\frac{\sin(\omega t_0+\theta)}{\omega}\,xP_x+\left(\frac{q{\cal{E}}'}{2}\frac{\sin(\omega t_0+\theta)}{\omega}\right)^2\,x^2,
\end{equation}
and similary for $y$. The term linear in $\Delta t$ in the expansion of $A_s$ about $t_0$ cancels the $-\Delta tE_0'(s)$ term, as it should but there is a remaining term quadratic in $\Delta t$, the bunching effect. This leaves
\begin{equation}
 -qA_s-\Delta tE_0'(s)=q{\cal{E}}\frac{\sin(\omega t_0+\theta)}{\omega}\left(1-\frac{\omega^2(\Delta t)^2}{2}\right)-\frac{r^2q}{4}\left({\cal{E}}''+\frac{\omega^2}{c^2}{\cal{E}}\right)\frac{\sin(\omega t_0+\theta)}{\omega}
\end{equation}
Notice the first term here and the first term in eqn.\,\ref{hamdt} depend only on the independent variable and not on the 6 dependent ones. Thus these do not affect the equations of motion and we ignore them. We have:
\begin{eqnarray}\nonumber
H_{\Delta t}&=&-\frac{q{\cal{E}}}{2}\omega^2T(\Delta t)^2+\frac{(\Delta E)^2}{2\beta^3\gamma^3mc^3}-\frac{r^2q}{4}\left({\cal{E}}''+\frac{\omega^2}{c^2}{\cal{E}}\right)T\\\nonumber
& &+\frac{P_x^2}{2P}-q{\cal{E}}'T\,\frac{xP_x}{2P}+\left(\frac{q{\cal{E}}'}{2}T\right)^2\,\frac{x^2}{2P} \\
& &+\frac{P_y^2}{2P}-q{\cal{E}}'T\,\frac{yP_y}{2P}+\left(\frac{q{\cal{E}}'}{2}T\right)^2\,\frac{y^2}{2P}
\end{eqnarray}
We defined here $T(s)=\sin[\omega t_0(s)+\theta]/\omega$ to clean up the notation a bit.

Finally, we wish to transform from $(\Delta t,-\Delta E)$ to $(z,P_z)=(-\beta c\Delta t,\Delta E/(\beta c))$. (The reason for the sign change is as follows: an early arrival implies $\Delta t<0$, but this means the particle is {\bf ahead} so $z>0$.) The generating function is 
\begin{equation} G=-\beta c\Delta tP_z\end{equation} 
(Check: $\frac{\partial G}{\partial\Delta t}=-\Delta E$, $\frac{\partial G}{\partial(P_z)}=z$.) The term to be added to the Hamiltonian is 
\[\frac{\partial G}{\partial s}=\frac{\beta'}{\beta}\,zP_z=\frac{\gamma'}{\beta^2\gamma^3}\,zP_z=\frac{q{\cal{E}}C}{\beta cP\gamma^2}\,zP_z,\] where $C\equiv\cos(\omega t_0+\theta)$.
\begin{eqnarray}\nonumber H_{z}&=&
  \frac{P_x^2}{2P}-q{\cal{E}}'T\,\frac{xP_x}{2P}+\left[\frac{1}{P}\left(\frac{q{\cal{E}'}T}{2}\right)^2-\frac{T}{2}\left(q{\cal{E}}''+\frac{\omega^2}{c^2}q{\cal{E}}\right)\right]\,\frac{x^2}{2}+\\\nonumber
&&\frac{P_y^2}{2P}-q{\cal{E}}'T\,\frac{yP_y}{2P}+\left[\frac{1}{P}\left(\frac{q{\cal{E}'}T}{2}\right)^2-\frac{T}{2}\left(q{\cal{E}}''+\frac{\omega^2}{c^2}q{\cal{E}}\right)\right]\,\frac{y^2}{2}+\\
&&\frac{P_z^2}{2\gamma^2P}+\frac{2q{\cal{E}}C}{\beta c}\,\frac{zP_z}{2\gamma^2P}-\frac{q{\cal{E}}}{\beta^2c^2}\omega^2T\,\frac{z^2}{2}\label{hamgen}
\end{eqnarray} 

\subsection{Hamiltonian 2}
Ryne\cite{rynerfgap1991} has a transformation that gets rid of the second derivative of the on-axis elecric field. It's complicated. At the same time he transforms away the adiabatic damping; it's a neat and didactic trick but not strictly necessary for computational purposes. It is simple to just use $P_{x,y,z}$ directly and then just rescale by final $P$ at the end. 

But there's an easy way to get rid of the second derivative: it turns out that the vector potential can be simplified if we use a different Gauge.

I propose the following function
\begin{equation}
 \Psi(x,y,s,t)=-\frac{\cal{E}'}{2}\frac{\sin(\omega t+\theta)}{\omega}\frac{x^2+y^2}{2}
\end{equation}
Add the gradient of this function to the previous vector potential (\ref{ax},\ref{ay},\ref{as}). This zeroes both $A_x$ and $A_y$, leaving
\begin{equation}
 A_s=-{\cal{E}}(s)\left(1-\frac{\omega^2}{c^2}\frac{x^2+y^2}{4}\right) \frac{\sin(\omega t+\theta)}{\omega}
\end{equation}
This is considerably simpler, but now there is a scalar potential:
\begin{equation}
\Phi=-\frac{\partial\Psi}{\partial t}={\cal{E}'}\cos(\omega t+\theta)\frac{x^2+y^2}{4}
\end{equation}
Now if we expand the Hamiltonian, we get a different result:
\begin{eqnarray}\nonumber H_{z}&=&
  \frac{P_x^2}{2P}+\frac{P_y^2}{2P}+\frac{q}{2\beta c}\left({\cal{E}'}C-{\cal{E}}S\frac{\omega\beta}{c}\right)\frac{r^2}{2}+\\
&&\frac{P_z^2}{2\gamma^2P}+\frac{2q{\cal{E}}C}{\beta c}\,\frac{zP_z}{2\gamma^2P}-\frac{q{\cal{E}}\omega S}{\beta^2c^2}\,\frac{z^2}{2}\label{hamgen2}
\end{eqnarray} 
($C\equiv\cos(\omega t_0(s)+\theta)$, $S\equiv\sin(\omega t_0(s)+\theta)$)

This is not only much simpler than eqn.\,\ref{hamgen} ($P_x$ and $P_y$ have their usual definitions, no transverse cross terms, no $\cal{E}''$), but has nice intuitive explanations for the individual terms. (1) The factor in parentheses is precisely the integrand of eqn.\,4 of a short note I wrote in 1985\cite{baarbun} explaining the derivation of the focal power of an RF gap, e.g.\ a buncher. (2) Taking the limit as $\omega\rightarrow 0$ reproduces precisely the Hamiltonian of the DC accelerator I derived in 2010\cite{baartman2010acc}. Note that in that case, ${\cal{E}'}=-\phi''$.

\subsection{Infinitesimal Transfer Matrix}
A convenient and useful way of representing the equations of motion 
through the optical element is the so-called infinitesimal transfer
matrix approach\cite{sacherer1970rms}. The infinitesimal transfer matrix $F(s)$
is defined as $(T-I)/ds$ where $T$ is the transfer matrix from
$s$ to $s+ds$ and $I$ is the identity matrix. With this definition
one has for individual particles
\begin{equation}
\label{force}
  \frac{d{\bf x}}{ds}=F{\bf x},\makebox[1in]{where}
  {\bf x}\equiv
 \left(\begin{array}{c} x\\P_x\\y\\P_y\\z\\P_z \end{array}\right).
\end{equation}
Beams of
particles are conveniently represented by the so-called $\sigma$-matrix;
the elements of which represent second order moments of the beam\cite{sacherer1970rms}.
The $\sigma$-matrix and the transfer matrix $M$ transform through the system
according to the equations
\begin{eqnarray}
 \frac{d\sigma}{ds}&=&F\sigma+\sigma F^T,\label{sig}\\
 \frac{dM}{ds}&=&FM.\label{space}
\end{eqnarray}
where $F^T$ is the transpose of $F$. 

Now that the Hamiltonian for linear motion (eqn.\,\ref{hamgen2}) has been
obtained, it is a simple matter to find the infinitesimal transfer
matrix $F$. Writing the equations of motion
($x'=\partial H/\partial P_x$, $P_x'=-\partial H/\partial x$, etc.) in the
form of Eqn.\,\ref{force}, the following $F$-matrix is found for the axially symmetric linear accelerator: 
\renewcommand{\arraystretch}{1.5}
\begin{equation} \label{itm}
F = \left( \begin{array}{cccccc}   
          0 & \frac{1}{P} & 0 & 0 & 0 & 0 \\
{\cal{A}}(s) & 0 & 0 & 0 & 0 & 0 \\
          0 & 0 & 0 & \frac{1}{P} & 0 & 0 \\
0 & 0 & {\cal{A}}(s) & 0 & 0 & 0 \\
0 & 0 & 0 & 0 & \frac{\beta'}{\beta} & \frac{1}{\gamma^2P} \\
          0 & 0 & 0 & 0 & {\cal{B}}(s) & -\frac{\beta'}{\beta}
\end{array}\right).
\end{equation} 
\renewcommand{\arraystretch}{1}
where we have defined:
\begin{eqnarray}{\cal{A}}(s)&=&\frac{-q}{2\beta c}\left({\cal{E}'}C-{\cal{E}}S\frac{\omega\beta}{c}\right),\\             {\cal{B}}(s)&=&\frac{q{\cal{E}}\omega S}{\beta^2c^2}.\end{eqnarray}

\subsection{Space Charge}
Space charge forces depend recursively upon the $\sigma$-matrix elements, and are simply added to the focusing elements $F_{2n,2m-1}|_{n,m=1,2,3}$ of the element's infinitesimal transfer matrix such as eqn.\,\ref{itm} above. This technique is used in the code {\tt TRANSOPTR}, as described by de~Jong\cite{dejong1983first}. The resulting equations can only be solved numerically. 

The given references\cite{sacherer1970rms,dejong1983first} treat space charge in the non-relativistic regime, so it was not obvious that {\tt TRANSOPTR} was correct in the relativistic regime. There are two effects that need to be considered to generalize the equations: the space charge magnetic field, and bunch length contraction. For detailed derivations, the interested reader is referred to the Ph.D.\ thesis of Fubiani\cite{fubiani2005controlled}. The first effect requires dividing the space charge force by $\gamma^2$. The second requires that the Carlson elliptic integrals' arguments be modified. The 3 integrals for the 3 major axes are
\[R_D(u,v,w) = \frac{3}{2} \int_0^\infty \frac{dt}{ (t+w)\,\sqrt{(t+u)(t+v)(t+w)}}\]
where $(u,v,w)$ are $(\sigma_{33},\sigma_{55},\sigma_{11})$ for the $x$-axis, $(\sigma_{55},\sigma_{11},\sigma_{33})$ for the $y$-axis, $(\sigma_{11},\sigma_{33},\sigma_{55})$ for the $z$-axis.\footnote{Note that this assumes the bunch axes are aligned with the beam direction $(s)$ and the chosen transverse orientation. If this is not the case, a rotation must be applied.} To be relativistically correct, $\sigma_{55}$ must be replaced by $\gamma^2\sigma_{55}$. (See Fubiani\cite{fubiani2005controlled}, Appendix\ J.) 

An interesting limit is the long bunch, since this can be approximated as a continuous beam with current $I$. First of all, it is clear that for this limit to apply, bunch length $\gg$ transverse size is not a necessary condition. Rather, $\gamma$ times bunch length $\gg$ transverse size. This means that for example a 1\,mm long by 1\,mm wide electron bunch is already well into the long-bunch regime with energy of 10\,MeV.

Secondly, in the long bunch regime, the Carlson integrals governing transverse space charge are $\propto (\gamma\sqrt\sigma_{55})^{-1}$, or the inverse bunch length augmented by the factor $\gamma$. For this reason, the constant governing space charge force in {\tt TRANSOPTR} in the unbunched case is divided by an extra factor of $\gamma$.

\section{Implementation into \tt TRANSOPTR}
In {\tt TRANSOPTR}, the momentum is dimensionless and in absence of acceleration corresponds to angles $(x',y',z')$. When acceleration occurs, angles are not scaled canonical momenta. The simplest implementation is to scale $(P_x,P_y,P_z)$ by initial total momentum $P_i$. Then after integration is complete, the momenta can be converted back into angles by multiplying by $P_i/P_f$. In this case, $F$ (eqn.\,\ref{itm}) becomes:
\renewcommand{\arraystretch}{1.5}
\begin{equation} \label{itm2}
F = \left( \begin{array}{cccccc}   
          0 & \frac{P_i}{P} & 0 & 0 & 0 & 0 \\
\frac{{\cal{A}}(s)}{P_i} & 0 & 0 & 0 & 0 & 0 \\
          0 & 0 & 0 & \frac{P_i}{P} & 0 & 0 \\
0 & 0 & \frac{{\cal{A}}(s)}{P_i} & 0 & 0 & 0 \\
0 & 0 & 0 & 0 & \frac{\beta'}{\beta} & \frac{P_i}{\gamma^2P} \\
          0 & 0 & 0 & 0 & \frac{{\cal{B}}(s)}{P_i} & -\frac{\beta'}{\beta}
\end{array}\right).
\end{equation} 
\renewcommand{\arraystretch}{1}

In order to find the $\sigma$-matrix at any point $s$ of a beamline, it is sufficient to calculate the transfer matrix to that point. This technique is not useful when space charge forces are non-negligible since then the optics themselves depend upon the $\sigma$-matrix. In the past space charge has been added to legacy codes such as {\tt TRANSPORT} by inserting a sufficient quantity of defocussing lenses whose strength depends upon the local beam size. This is inefficient at best: it is the crudest sort of integration. The best way to do this is as implemented in {\tt TRANSOPTR}: the exact Sacherer 6D envelope equations (\ref{sig}) are integrated with a higher order integrator such as the Runge-Kutta.

In many situations, one needs only the $\sigma$-matrix and so it is sufficient to solve the $36$ equations of eqns.\,\ref{sig}.\footnote{In fact, since this matrix is symmetric, one can get away with solving only $21$ equations. One can whittle this down further for uncoupled cases; in fact $4$ first order or $2$ second order equations are often all that is needed for a straight beamline with no coupling and no dispersion. These are of course the Kapchinsky-Vladimirsky equations.} However, in general one would like to have the ability to fit certain optical characteristics, such as point-to-point imaging. It is therefore often useful to also know the transfer matrix to any point. This is in fact the {\bf incoherent} transfer matrix. So in order to cover all cases, {\tt TRANSOPTR} solves both eqns.\ \ref{sig} and \ref{space}. This comprises $72$ equations. Internally, the code uses a single $12$ by $6$ matrix.

Without time-varying fields, the total momentum of the reference particle is always known at any $s$. In a linac, this is no longer true. Thus, the implementation needs to track not only the transfer and $\sigma$ matrix elements, but also time and energy. These can be thought of as the 5th and 6th ``zero-order'' coordinates, with respect to which the fifth and sixth phase space coordinates are measured. Analogously, in the most general cases, the first through fourth ``zero-order'' coordinates may also not be known. This could happen for example if given only the field map of a dipole magnet. Thus there could be 6 ``zero-order'' coordinates as a function of $s$, and the usual 6 phase space coordinates are differentials with respect to these.

I therefore have added a row to the 72-element matrix, making it dimensioned 13-by-6, and there are 78 ODEs. This would allow at a later date adding elements whose reference trajectory is not known a priori.

\subsection{Code}
Here follows the f77 subroutine that takes the 13-by-6 matrix {\tt SX} and calculates the 13-by-6 matrix {\tt DSX}, which are the derivatives of {\tt SX} with respect to {\tt Z}=$s$. Note that {\tt SX(13,5)} is not time $t$ directly but is $ct$, and {\tt SX(13,6)} is not energy directly but is $\gamma-1$. The $F$ matrix is in fact {\tt F} directly.

\begin{verbatim}
      SUBROUTINE SCLINAC(Z,SX,DSX)
      COMMON/PRINT/IPRINT,IQ1,JQ1,IQ2,JQ2,IQ3,JQ3,IQ4,JQ4
      COMMON/VELEM/VELM(21,9999),NELM
      DIMENSION SX(13,6),DSX(13,6)
      DIMENSION TEMP(6,6),ROT3(3,3),ROT6(6,6),SIG(3,3)
      DIMENSION WRK(3),EVAL(3),RF(6,6),FT(6,6)
      LOGICAL STRATE
      common/splind/klob,kloe
      COMMON/FMATRX/F(6,6)
      COMMON/SCPARM /KSC,ISC
      COMMON/CALLS/NSC
      COMMON/MOM/P,BRHO,pMASS,ENERGK,GSQ,ENERGKi,charge,bncharge
      COMMON/axezrf/omoc,phase,etot,
     $     ZTBLE(9999),EZTBL(9999),SPCFE(9999),zse,iez,iaxezrf
      COMMON/axez2/pratio
      REAL KSC
      do i=1,4
      dsx(13,i)=0.
      enddo
      NSC=NSC+1
      pratio=1.
      gamm1=sx(13,6) ! gamm1=gamma-1
      energk=gamm1*PMASS
         if (energk.le.0.)then
            write(6,*)'*********BEAM DECELERATED TO A STOP'
            write(6,*)'Check parameter values or limits.'
            call errout
            endif
c call spline routine to find the on-axis electric field and derivatives
          call spling(kloe,ZTBLE,EZTBL,SPCFE,IEZ,z-zse,EZ,EZp,EZpp)
          ccc=cos(sx(13,5)*omoc+phase)
          sss=sin(sx(13,5)*omoc+phase)
c omoc is omega/c
          dsx(13,6)=EZ*ccc/PMASS/100. ! denergy/ds ; EZ is in MeV/m, already contains CHARGE
          gamma=1.+gamm1
      ETA=SQRT(gamm1*(gamm1+2.))
      BRHO=ETA*PMASS/CHARGE*3.3356397
      BRHO=BRHO*.001
      P=ETA*PMASS !this P is actually Pc
      gsq=GAMMA**2
      BETA=eta/gamma
          dsx(13,5)=1./beta ! dctime/ds=1./beta
      gamm1old=etot/pmass
      ETAold=SQRT(gamm1old*(gamm1old+2.))
      Pold=ETAold*PMASS 
      pratio=etaold/eta
c re-calculate space charge parameter as energy changes.
c (898755.2 is 1/(4pi epsilon_0) ~ c^2/1.e7 in some units)
      KSC=BNCHARGE*CHARGE*898755.2/(ETA**2*PMASS)/pratio
            F(1,2)=pratio
            F(3,4)=pratio
            F(2,1)=(EZ*sss*omoc-EZP*ccc/beta)/(200.*Pold)
            F(4,3)=F(2,1)
c bpob = beta'/beta
            bpob=dsx(13,6)/(gamma*eta**2)
            F(5,5)=bpob
            F(5,6)=pratio/gsq
            F(6,5)=EZ*sss*omoc/(100.*Pold*beta**2)
            F(6,6)=-bpob
      IF (KSC.NE.0.)THEN
c *******************************************************************
c START SPACE CHARGE SECTION:
c *******************************************************************
      STRATE=.TRUE.
c SIG is real-space alone (3D)
      DO 5 I=1,2
         DO 5 J=1,2
 5    SIG(I,J)=SX(2*I-1,2*J-1)
      gamma=sqrt(gsq)
C QUICK AND DIRTY Lorentz transform.
      SIG(1,3)=GAMMA*SX(1,5)
      SIG(2,3)=GAMMA*SX(3,5)
      SIG(3,1)=SIG(1,3)
      SIG(3,2)=SIG(2,3)
      SIG(3,3)=GSQ  *SX(5,5)
c      write(36,*)'gsq',gsq,sig(3,3)
c *******************************************************************
c Test whether bunch axes are along coordinate axes
c *******************************************************************
      IF (SIG(1,2).NE.0..OR.
     ,   SIG(1,3).NE.0..OR.
     ,   SIG(2,3).NE.0.)THEN
           CALL PRESET(ROT6,36,0.)
           STRATE=.FALSE.
c Find rotation required
      call jacobi(sig,3,3,eval,rot3,nrot)
c Fill 6D rotation matrix
      DO 55 I=1,3
         DO 55 J=1,3
            ROT6(2*I-1,2*J-1)=ROT3(I,J)
 55         ROT6(2*I,2*J)=ROT3(I,J)
      ELSE
c *******************************************************************
c Bunch is along axes
c *******************************************************************
           DO 52 I=1,3
 52        EVAL(I)=SIG(I,I)
      ENDIF
      do i=1,3
       if (eval(i).le.1.e-6)then
         eval(i)=1.e-6
         endif
         enddo
c RD is the Carlson elliptic integral R_D.
c Scale the arguments. Carlson routine is better if args are order 1. 
         scalrd=eval(1)
         if(eval(2).gt.scalrd)scalrd=eval(2)
         if(eval(3).gt.scalrd)scalrd=eval(3)
         rscalrd=scalrd**(-1.5)
      EL1=rscalrd*RD(EVAL(2)/scalrd,EVAL(3)/scalrd,EVAL(1)/scalrd)
      EL2=rscalrd*RD(EVAL(3)/scalrd,EVAL(1)/scalrd,EVAL(2)/scalrd)
      EL3=rscalrd*RD(EVAL(1)/scalrd,EVAL(2)/scalrd,EVAL(3)/scalrd)
      DO 70 I=1,6
      DO 70 J=1,6
 70   RF(I,J)=0.
      RF(2,1)=KSC*EL1
      RF(4,3)=KSC*EL2
      RF(6,5)=KSC*EL3*gsq
c *******************************************************************
c     no rotation required
c *******************************************************************
 60   IF (STRATE)THEN
         DO 40 I=1,6
         DO 40 J=1,6
 40      FT(I,J)=F(I,J)+RF(I,J)
      ELSE   
c *******************************************************************
c     rotate to lab frame
c *******************************************************************
      DO 15 I=1,6
         DO 15 J=1,6
            TEMP(I,J)=0.
         DO 15 M=1,6
15    TEMP(I,J)=TEMP(I,J)+RF(I,M)*ROT6(J,M)
      DO 20 I=1,6
         DO 20 J=1,6
            FT(I,J)=F(I,J)
            DO 20 M=1,6
20    FT(I,J)=FT(I,J)+ROT6(I,M)*TEMP(M,J)
      ENDIF
c *******************************************************************
c END SPACE CHARGE SECTION
c *******************************************************************
c START NO SPACE CHARGE SECTION
c *******************************************************************
      ELSE                                     !ie. KSC=0.
      DO 23 I=1,6
         DO 23 J=1,6
23          FT(I,J)=F(I,J)
      ENDIF 
c *******************************************************************
c END NO SPACE CHARGE SECTION
c *******************************************************************
c MATRIX OPERATIONS TO OBTAIN DIFF. EQUATIONS
c *******************************************************************
      DO 30 I=1,6
         DO 30 J=1,I
            DSX(I,J)=0.
            DO 25 M=1,6
25          DSX(I,J)=DSX(I,J)+FT(I,M)*SX(M,J)+SX(I,M)*FT(J,M)
30    DSX(J,I)=DSX(I,J)
      DO 26 I=7,12
         DO 26 J=1,6
            DSX(I,J)=0.
            DO 26 M=1,6
26          DSX(I,J)=DSX(I,J)+FT(I-6,M)*SX(M+6,J)
      RETURN
      END
\end{verbatim}

\section{Example}
The TRIUMF injector electron linac, EINJ, takes bunches from energy of 300\,keV to $\sim10$\,MeV if properly phased and the peak gradient is $20$\,MV/m. Below is example for phase $\theta=0$ at the start of the calculation.

\includegraphics[width=\textwidth]{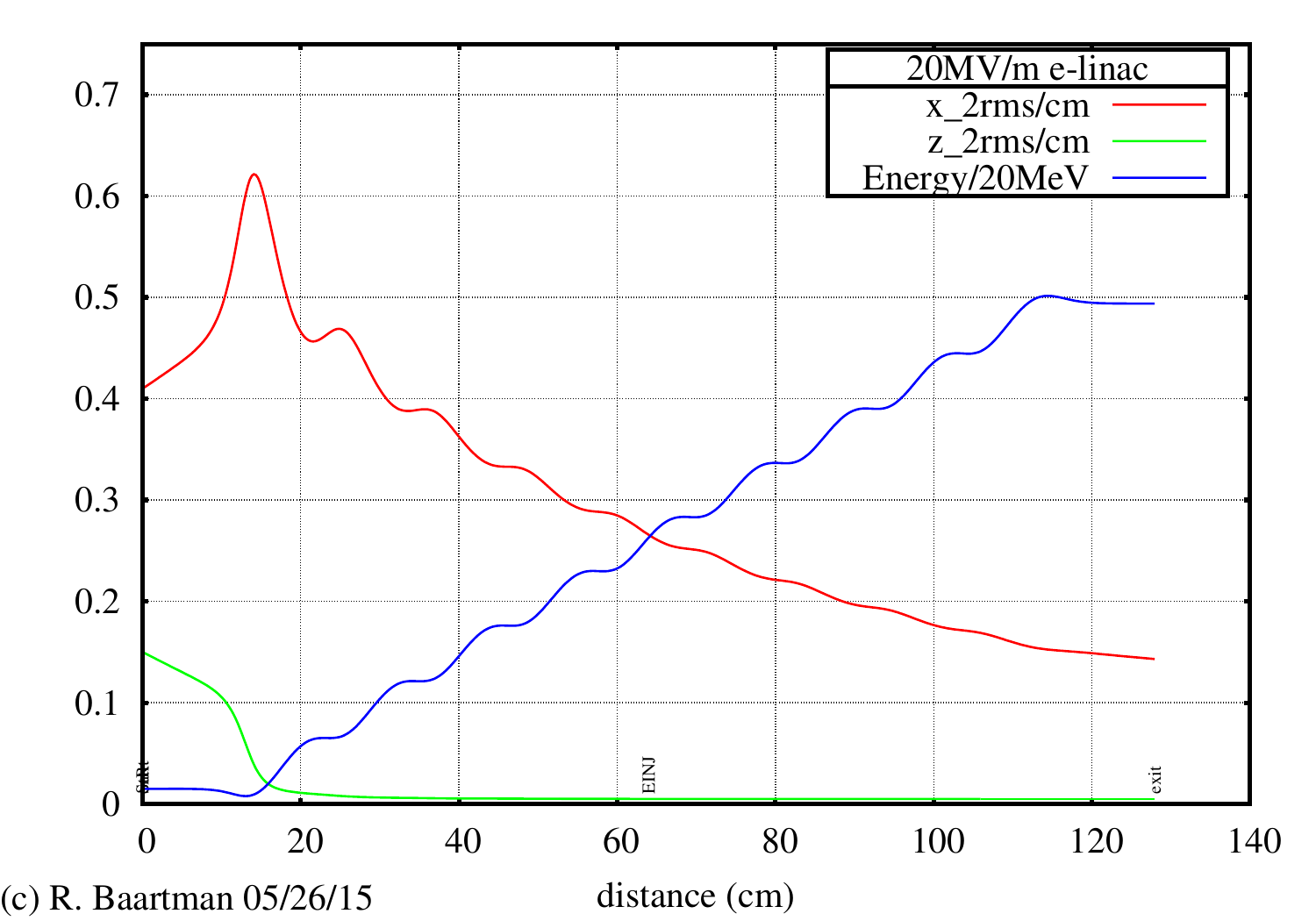}

Red is the 2rms transverse size, and green is the 2rms longitudinal (bunch length). The input bunch parameters are somewhat arbitrary, roughly the condition for a minimum beam size at exit. This particular case has zero bunch charge.

In this second example, {\tt TRANSOPTR} is instructed to fit the $65$ matrix element to zero. This makes energy insensitive to input phase, thus finding the peak energy gain phase. This phase turns out to be $\theta=-15.46^\circ$.

\includegraphics[width=\textwidth]{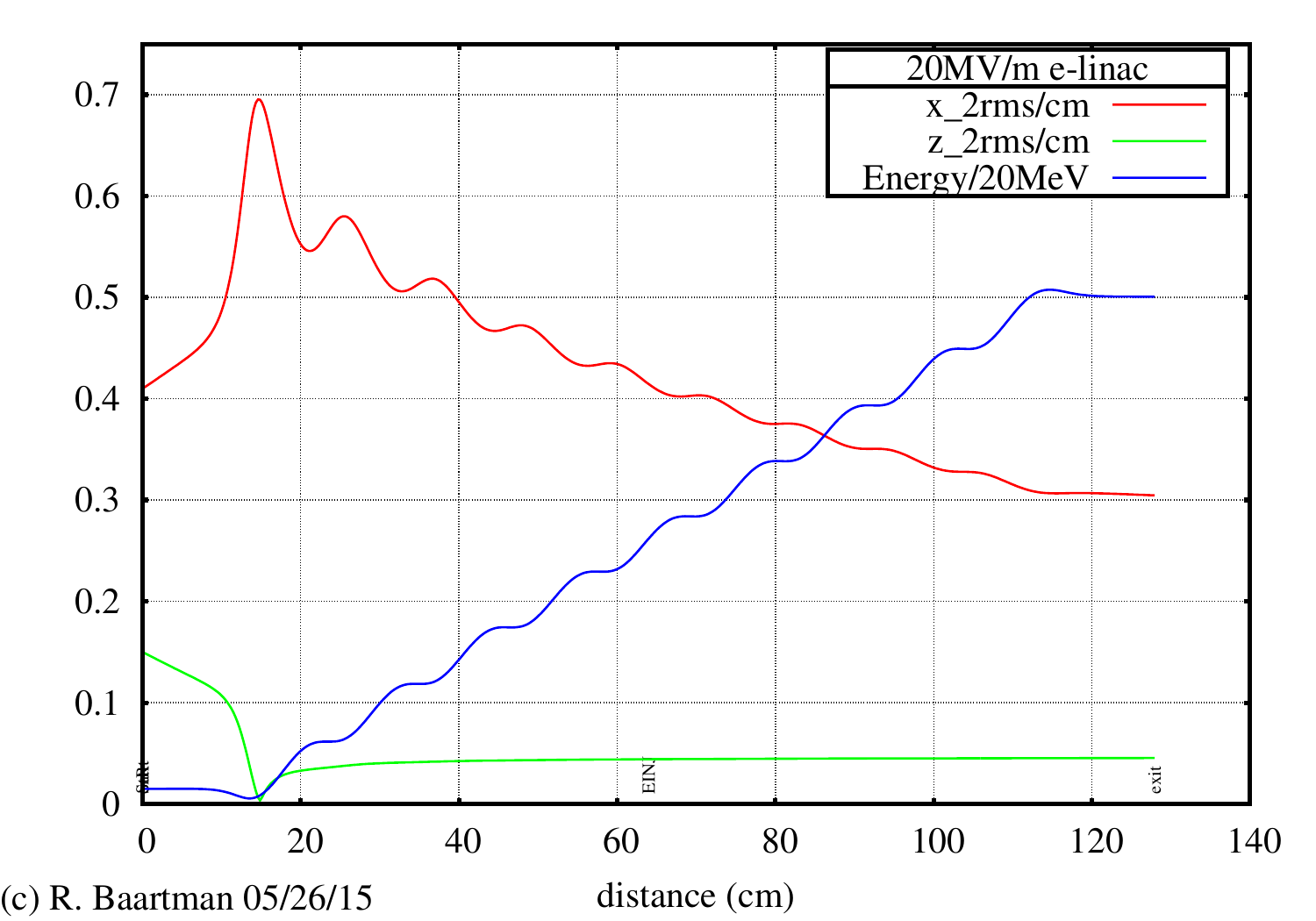}

In the third example, bunch charge has been raised to $30$\,pC.

\includegraphics[width=\textwidth]{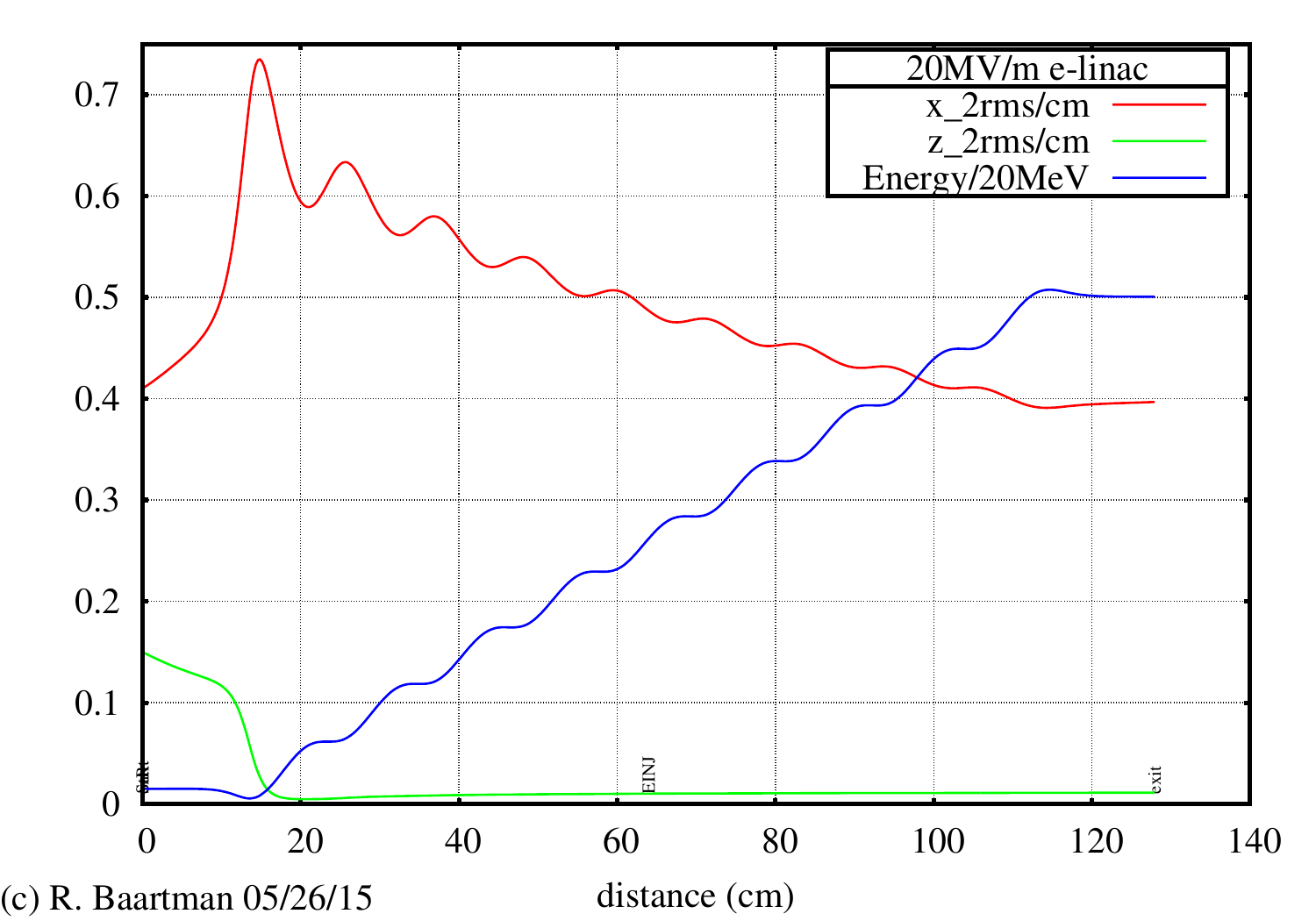}

\subsection{Timing}
Each calculation above takes roughly 400 Runge-Kutta steps for 2400 calls to the {\tt SCLINAC} routine. This gives 5-figure accuracy to the transfer matrix and the $\sigma$-matrix, and is easily enough for describing reality considering that the on-axis field is only known to 2 or 3 significant figures. The extra accuracy is useful, however for fitting matrix or beam matching, which is done with a downhill simplex method, or simulated annealing for cases of more than 3 fitting parameters.

On my unremarkable, circa 2006 Intel desktop, each run through the linac takes about 17 milliseconds with zero bunch charge and 25 milliseconds with space charge. The difference is due to the Carlson elliptic integrals needed for the space charge case.

On a typical optics matching case, one varies 2 solenoids, the buncher amplitude, and the linac phase, to minimize the bunch size and energy spread at the linac output. A calculation with such a fit requires typically a half million total calls to {\tt SC} (the space charge routine for no-linac case) and {\tt SCLINAC}, and so takes about 5 seconds CPU time. The result is shown below. The bunch charge is $15$\,pC. 

Rather extravagantly, each calculation starts from the cathode whereas it would have been more efficient to store the beam parameter set at the buncher entrance and start it from there. The DC acceleration to 300 keV from the cathode is described in reference \cite{baartman2010acc}.

The Buncher itself, located at $s=85$\,cm, is calculated as just another linac, phased to give no energy gain.

\includegraphics[width=\textwidth]{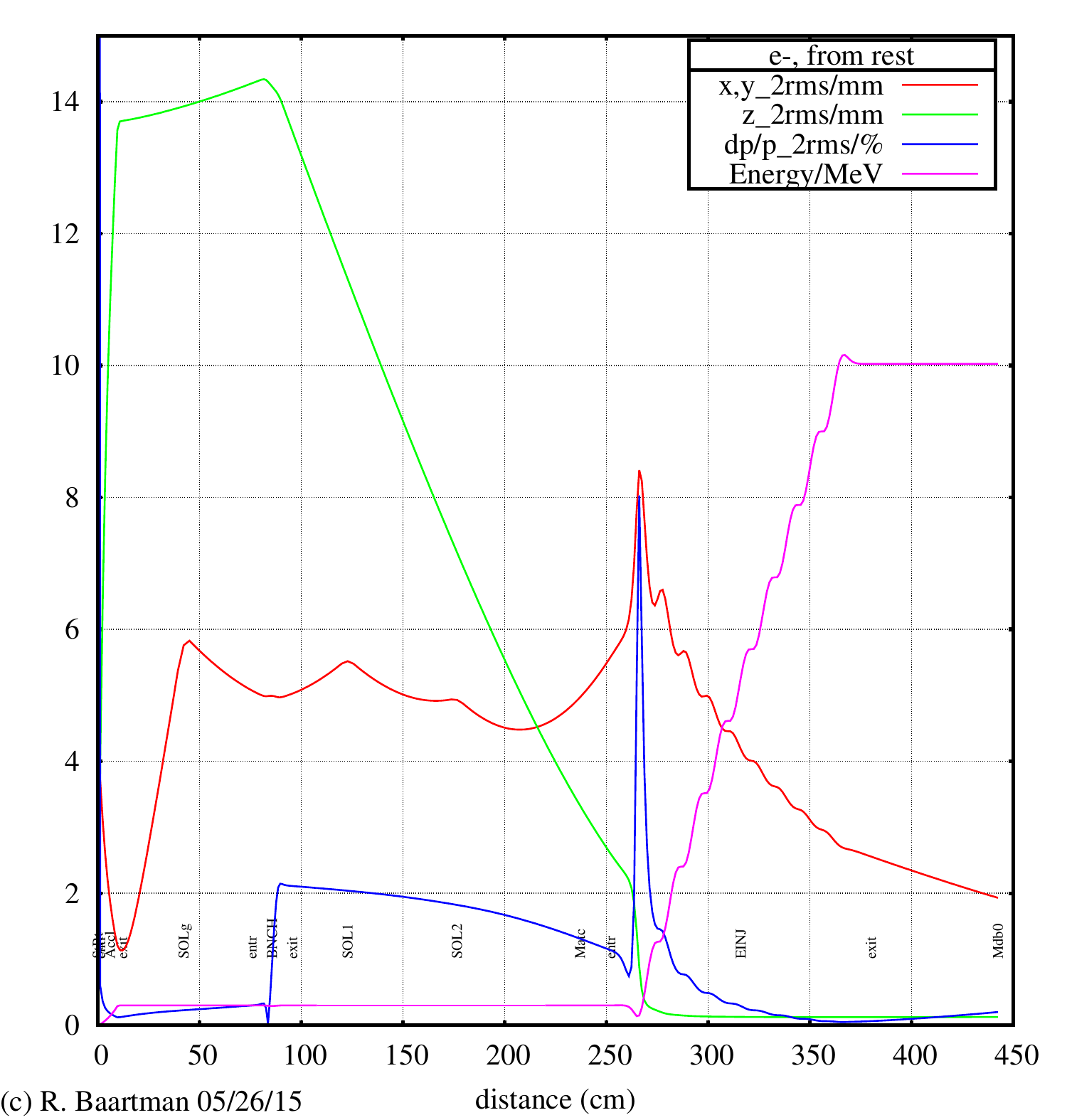}

The code describing the whole transport from cathode to linac exit, is shown below.

\begin{verbatim}
       SUBROUTINE TSYSTEM
      COMMON /BLOC1/c0,Ebun,phbun,c1,c2,phase,grad,birk
      COMMON/MOM/P,BRHO,pMASS,ENERGK,GSQ,ENERGKi,charge,current
      COMMON/SS/SX(13,6)
      nss=2
c      sollen=6.32 this was with old clamps
      sollen=5.64
      sollenrot=7.29
      solt=c0*140.0/10000.
      sol1=c1*116.9/10000.
      sol2=c2*116.9/10000.
c following are from autocad dwg.
      atgunsol=42.193
      atdb=56.9
      atrfgap=85.642
      atdb2=105.7
      atsol1=122.777
      atsol2=176.577
      atmatch=237.
      sap=2.5/2.*2.54
      sss=sollen/nss
c since actually a solenoid's aberration is unrelated to its length,
      wt=ww/sqrt(float(nss))
      acclen=12.0
      call axez(59,121,-100.,acclen,10,1,2)
      call drs(atgunsol-acclen-sollen/2.)
      rho=100.*brho/solt
      th= sss/rho/2.*57.29578
      do i=1,nss
         if(i.eq.(nss+1)/2)then
      call solenoidn(solt,sss,sap,wt,'SOLg')
         else
      call solenoidn(solt,sss,sap,wt,'   .')
         endif
      call rotate(-th,'   .')
      enddo
      th= sollenrot/rho/2.*57.29578
      call rotate(th,'   .')
      if(iprint.ne.0)write(6,*)'Sol.B/1gauss,rot.= ',solt*10000.,th*nss
      call drs(atdb-atgunsol-sollen/2.)
      call drs(atrfgap-atdb-10.) !-ccy
      call linacn(62,21,Ebun,20.,1.3e9,phbun,20,'BNCH') !BuncherEH
      sollen=5.85
      sollenrot=7.37
      sss=sollen/nss
      call drs(atdb2-atrfgap-10.)
      call drs(atsol1-atdb2-sollen/2.)
      rho=100.*brho/sol1
      th= sss/rho/2.*57.29578
      do i=1,nss
         if(i.eq.(nss+1)/2)then
      call solenoidn(sol1,sss,sap,wt,'SOL1')
         else
      call solenoidn(sol1,sss,sap,wt,'   .')
         endif
      call rotate(-th,'   .')
      enddo
      th= sollenrot/rho/2.*57.29578
      call rotate(th,'   .')
      if(iprint.ne.0)write(6,*)'Sol.B/1gauss,rot.= ',sol1*10000.,th*nss
      call drs(atsol2-atsol1-sollen)
      rho=100.*brho/sol2
      th= sss/rho/2.*57.29578
      do i=1,nss
         if(i.eq.(nss+1)/2)then
      call solenoidn(sol2,sss,sap,wt,'SOL2')
         else
      call solenoidn(sol2,sss,sap,wt,'   .')
         endif
      call rotate(-th,'   .')
      enddo
      th= sollenrot/rho/2.*57.29578
      call rotate(th,'   .')
      if(iprint.ne.0)write(6,*)'Sol.B/1gauss,rot.= ',sol2*10000.,th*nss
      d5=74.96-sollen/2.
      call drs(atmatch-atsol2-sollen/2.)
      call drift(0.,'Matc')
      call drs(15.-1.023) 
      call fit(1,1,1,0.,1.,1)
      call fit(1,5,5,0.4,8.,1)
      call drift(1.023,'   .')
      sx(13,5)=0.
      call linacn(71,447,grad,127.953,1.3e9,phase,100,'EINJ')
      call drift(1.023,'   .')
      call fit(1,1,1,0.,4.,1)
      call fit(1,2,2,0.,1000.,1)
      call fit(1,5,5,0.,70.,1)
      call fit(1,6,6,0.,1000.,1)
      enrg=sx(13,6)*0.511
c we don't want output energy to be too far off.
      call fitarb(10.02,enrg,100.,1)
      call drs(61.37) 
      call drift(0.,'Mdb0')
      call vective(1)
      return
      end
\end{verbatim}

\bibliographystyle{unsrt} 
\bibliography{Baartman,BaartmanDN,Others}
\end{document}